\documentstyle[12pt,epsfig]{article}
\def \ins#1#2#3#4#5#6#7 {
  \begin{figure}[#6]
    \hskip #5
    \special{em:graph {./pcx/#2} }
    \vskip #4
    \caption{#7}
    \label{#1}
  \end{figure}}

\begin{document}

\begin{center}
{\Large \bf
Estimation of admixture of twelve quark bag state in  ${}^4He$
nucleus}
\end{center}

\begin{center}
A.M. Mosallem \footnote{Math. \& Theor. Phys. Dept., NRC,
AEA, Cairo, Egypt.}, V.V.Uzhinskii
\end{center}

\begin{center}
Joint Institute for Nuclear Research, \\
Laboratory of Information Technologies
\end{center}
\noindent
The $p{}^4He$ elastic scattering at the energy range from
0.695 to 393 GeV is analyzed in the framework of the Glauber
theory. The Glauber amplitudes were evaluated using isospin-averaged
nucleon-nucleon amplitudes and the ${}^4He$ wave function as a
superposition of the Gaussian functions. The values of the
calculated differential cross sections usually exceed the experimental
ones.  In order to overcome the discrepancy, it is assumed following to
the paper by L.G. Dakno and N.N. Nikolaev (Nucl. Phys. {\bf A436}
(1985) 653) that the ground state wave function of ${}^4He$ has an
admixture of a twelve quark bag. Neglecting all transition amplitudes,
the proton - 12q bag scattering amplitude was chosen in a simple
gaussian form. The inclusion of the 12q bag leads to decreasing
the $p{}^4He$ differential cross section and to a shift of the dip
position to a large values of $t$ what is needed for a successful
description of the experimental data. While fitting the data it is
found that the weight of the 12q bag state in the ground state of the
${}^4He$ nucleus is $\sim 10.5$\%, $\sigma ^{tot}_{p-12q}\sim 34$ mb,
and the slope parameter of the $p-12q$ bag elastic scattering is $\sim$
23 (GeV/c)$^{-2}$.  Inelastic shadowing is not taken into account at
the calculations.

\newpage
\setcounter{page}{1}
\section*{Introduction}
The study of structure of light nuclei such as ${}^4He,{}^6He,{}^{11}Li$
and so on is very popular now. There is a
big progress in understanding the structure of light exotic
nuclei ${}^6He,{}^{11}Li,...~$. The deuteron structure is a
subject of continuous discussions. Only few hypotheses about the
structure of ${}^4He$ exist now.  In paper by L.G. Dakno and N.N.
Nikolaev \cite{Nik-Dakhno} it was assumed and shown that $12\%$
admixture of twelve quark bag configuration in the ground state wave
function of the ${}^4He$ nucleus allows one to understand the
irregularities of proton -${}^4He$ elastic scattering at high energies.
We believe that the hypothesis will permit to describe other
reactions -- $d+{}^4He, {}^4He+{}^4He,{}^4He+C,...$ etc. The
matter is the carbon and oxygen nuclei are considered as strong
clustering nuclei, consisting of $\alpha$ particles. So, the
peculiarities of the ${}^4He$ nucleus can manifest themselves in the
structure of ${}^{12}C$ and ${}^{16}O$ nuclei.

To show these, one needs to calculate elastic and inelastic
scattering of ${}^4He$ on different nuclei. The Glauber diffraction
theory \cite{Glauber} of multiple scattering processes has been
generally accepted as a suitable framework for such calculations. But
it was recognized many years ago that the model predictions have been
far from being perfect even for the hadron-nucleus scattering process.

Many authors believe that it is due to inelastic screening,
and many attempts have been made to take them into account
\cite{Abers,Gribov, Pump,Alberi,Harri,Quigg}. According to different
calculations, the inelastic screening corrections to the total
hadron-nucleus cross sections are at the level of 2--5 \%.  It is not
enough to describe the $p{}^4He$ scattering. Inclusion of the
corrections into calculations of the $p{}^4He$ elastic scattering
leads to a shift of the first diffraction minimum to low values of the
momentum transfer, $t$. But a good description of the cross sections
demands shift of the minimum to large values of $t$. In order to
solve the problem at the first step of our study, we will omit the
corrections.

The content of the paper is as follows: Sec. 1 describes
calculation of the ${}^4He$ form factor with different parametrizations
of the ground state wave function. Sec. 2 gives calculations of the
$p{}^4He$ elastic scattering amplitude and differential cross section
with these parametrizations. In Sec. 3 we include the twelve
quark bag admixture and fit the parameters of the twelve quark bag using
experimental data. In the last Sec. we summarize our results.

\section{Form-factor of ${}^4$He}
The main characteristic properly of a nucleus is a nuclear
form-factor.
\begin{equation}                                 
F(\vec q)=\int{e^{i \vec q \cdot {\vec r}_1}\left|{\psi}({\vec
r}_1,\ldots,{\vec r}_A)\right|}^{2}\prod_{i=1}^{A}d^{3}r_i,
\label{eq1}
\end{equation}                               
\noindent where $\psi$ is the wave function of a nucleus in the ground
state, A -- a mass number of the nucleus, $\vec r_1,~ \vec r_2,~
\ldots$ -- radius vectors of nuclear nucleons, $\vec q$ - momentum
transfers. It is very often assumed in Glauber calculations that the
square module of $\psi$ can be represented as
\begin{equation}                                
{\left|{\psi}({\vec r}_1,\ldots,{\vec r}_A)\right|}^{2}= (2\pi)^3
\rho_c \delta\left(\sum_{i=1}^{A}{\vec r}_i\right)\prod_{i=1}^{A}
\varphi(\vec r_i).
\label{eq2}
\end{equation}                                  
\noindent The $\delta$-function is introduced in order to satisfy the
obvious condition
\begin{equation}                                
\left(\sum_{i=1}^{A}{\vec r}_i\right)=0.
\label{eq3}
\end{equation}                                  
For the ${}^4$He nucleus in the paper \cite{Nik-Dakhno} the following
parametrizations of $\varphi(\vec r)$ were proposed:
\begin{eqnarray}
(A)&& \varphi({\vec r})= exp[-{\vec r}^{2}/R_{1}^{2}],
\nonumber \\                                    
(B)&& \varphi({\vec r})= exp[-{\vec r}^{2}/R_{1}^{2}]-D_{1}exp[-{\vec
r}^{2}/R_{2}^{2}],
\nonumber \\                                    
(C)&& \varphi({\vec r})= (exp[-{\vec r}^{2}/2R_{1}^{2}]-D_{1}exp[-{\vec
r}^{2}/2R_{2}^{2}])^{2},
\nonumber \\                                    
(D)&&\varphi({\vec r})= exp[-{\vec r}^{2}/R_{1}^{2}]+D_{1}exp[-{\vec
r}^{2}/R_{2}^{2}]-(1+D_{1}-D_{2}^{2})exp[-{\vec
r}^{2}/R_{3}^{2}]. \nonumber
\nonumber
\end{eqnarray}
The parameters are given in Table 1.
\begin{table} [h]                         
\centering
\caption{Values of used parameters (from \protect{\cite{Nik-Dakhno}})}
\begin{tabular}{|c|c|c|c|c|c|} \hline
         &  $R_1^2$  &  $R_2^2$  &  $R_3^2$  & $D_1$ & $D_2 $ \\ 
         &$(GeV/c)^{-2}$&$(GeV/c)^{-2}$&$(GeV/c)^{-2}$& &  \\ \hline
    $A$  &  51.01   &          &          &       &         \\ \hline
    $B$  &  48.07   &  3.67    &          &  1.0  &         \\ \hline
    $C$  &  47.29   &   1.6    &          &  1.6  &         \\ \hline
    $D$  &  62.06   &   19.0   &   10.13  &  3.79 & 0.31     \\ \hline
\end{tabular} \end{table}                     

\noindent We will use a general form for the function $\varphi$ as
\begin{equation}                                   
\phi({\vec r})= \sum_{i=1}^{N}C_{i}e^{{\vec r}^2/R^{2}_{i}}.
\label{eq4}
\end{equation}                                     

In Eq. (\ref{eq2}) $\rho_c$ is the normalization constant
\noindent determined from the condition
\begin{equation}                                 
\int \left|\psi(\vec r_1,\ldots,\vec r_A) \right|^2 \prod_{i=1}^A
d^3 r_i =1.
\label{eq5}
\end{equation}                                   
Substituting Eq. (\ref{eq2}) in the normalization condition
(\ref{eq5}), we have
\begin{equation}                                  
\rho_c (2\pi)^3 \int \delta \left( \sum_{i=1}^{4} {\vec r}_i \right)~
\prod_{i=1}^{4} \varphi (\vec r_i) d^3 r_i=1.
\label{eq6}
\end{equation}                                    
Using the following representation of the $\delta$
function
$$                                                 
\delta \left(\sum_{i=1}^{4}{\vec r}_i\right)= \frac {1}{(2\pi)^3}\int
d^3\alpha~e^{i \vec \alpha \cdot \left( \sum_{i=1}^{4}\vec
r_i \right) },
$$
the Eq. (\ref{eq6}) can be re-written as
\begin{equation}                                  
\rho_c\int ( 2 \pi)^3~\left( \frac{1}{2\pi} \right)^3
d^3{\alpha}~e^{i \vec \alpha \cdot \left(\sum_{i=1}^{4}{\vec
r}_i\right)} ~\prod_{i=1}^{4}{\varphi}({\vec r}_i)d^3r_i=1
\label{eq7}
\end{equation}                                    
\noindent Then
\begin{eqnarray}                                  
{\rho_c}^{-1}&=& \int d^3{\alpha}~\prod_{i=1}^{4} e^{i{\vec
\alpha}.{\vec r}_i} ~{\varphi}({\vec r}_i)d^3r_i  
   = \int d^3{\alpha}~\prod_{i=1}^{4}e^{i{\vec \alpha}.{\vec
r}_i} ~\left(\sum_{j=1}^{N}C_{j}e^{{\vec
r}_{i}^{2}/R^{2}_{j}}\right)d^3r_i.
\label{eq8}
\end{eqnarray}                                    
Integration with respect to $r_i$ gives
\begin{eqnarray}                                   
\rho_c^{-1}&=&\int d^3 \alpha \prod_{i=1}^4 \sum_{j=1}^N
C_j \left(\pi R_j^2\right)^{3/2} e^{-\alpha ^2 R_j^2/4} \nonumber
  =\int d^3 \alpha \left(\sum_{j=1}^N C_j \left(\pi
R_j^2 \right)^{3/2} e^{-\alpha^2 R_j^2/4}\right)^4
\nonumber\\                                        
    &=&\int d^3 \alpha
\sum_{i_1,i_2,i_3,i_4}^N C_{i_1} C_{i_2} C_{i_3} C_{i_4}
\left(\prod_{s=1}^4 \left( \pi R_{i_s}^2 \right)^{3/2}\right)
e^{-\frac{\alpha^2}{4}\left(\sum_{j=1}^4 R_{i_j}^2\right)},
\label{eq9}
\end{eqnarray}                                    
and final integration with respect to $\alpha$ yields
\begin{equation}                                  
\rho_c^{-1}=\sum_{i_1,i_2,i_3,i_4}^N C_{i_1} C_{i_2} C_{i_3}
C_{i_4} \left(\prod_{s=1}^4\left(\pi R_{i_s}^2\right)^{3/2}\right)
\left(\frac{4\pi}{\sum_{j=1}^4 R_{i_j}^2}\right)^{3/2}.
\label{eq10}
\end{equation}                                    

The one-particle density function is determined as
\begin{equation}                                  
\rho (\vec r) = \int {\left|\psi(\vec r,\vec r_2,\vec r_3,\vec
r_4)\right|}^{2} d^3 r_2 d^3 r_3 d^3 r_4,
\label{eq11}
\end{equation}                                    
and can be calculated in an analogous way.
The functions $\rho (\vec r)$ corresponding to the parametrizations ($A$
-- $D$) of the wave function are shown in Fig. 1. All densities  are
close to each other at large values of $r$, and they are
different in the nucleus center. So, the parametrizations take
various short range $NN$ correlations into account.
\begin{figure}[cbth]
\begin{center}
\psfig{file=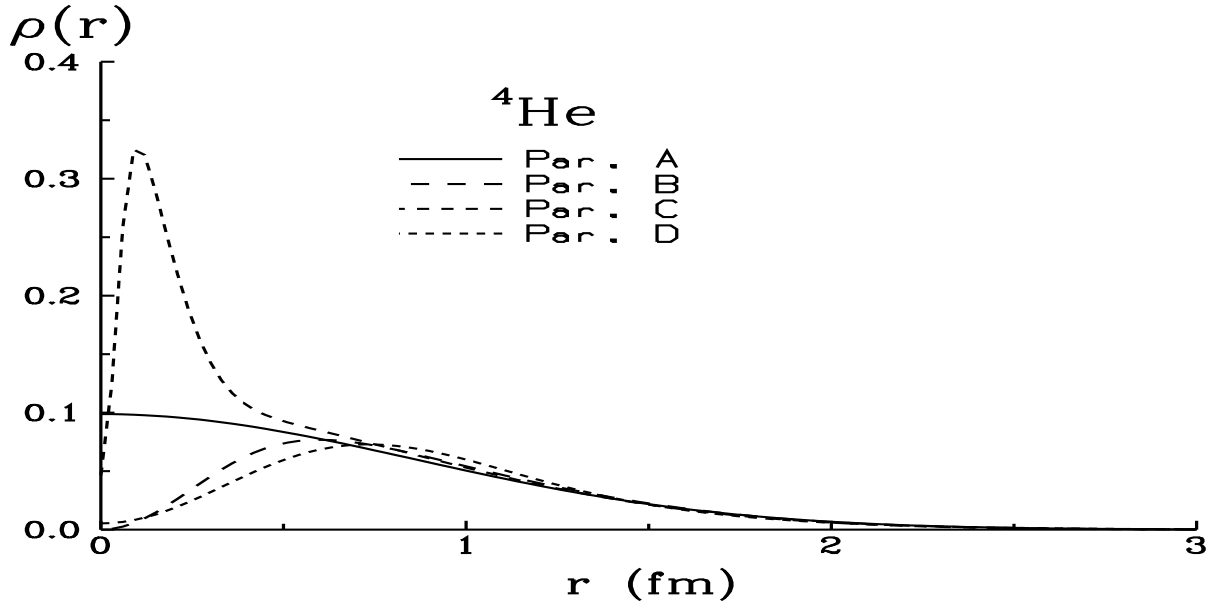,width=160mm,height=80mm,angle=0}
\caption{The one-particle density of the ${}^4He$}
\end{center}
\end{figure}

Performing nearly the same
calculations, we have the following expression for the form factor:
\begin{eqnarray}                            
F(\vec q)&=& \int e^{i \vec q \cdot {\vec r}_1}(2\pi)^3 \rho_c
\delta\left(\sum_{i=1}^{4}{\vec r}_i\right)\prod_{i=1}^{4}
\varphi(\vec r_i) d^{3}r_i
\nonumber \\                               
&=& (2\pi)^3 \rho_c \int \frac{d^3 \alpha}{(2\pi)^3} e^{i \vec q \cdot
\vec r_1} e^{i \vec \alpha \cdot \left(\sum_{i=1}^{4} \vec r_i\right)}
\prod_{i=1}^{4}\left(\sum_{j=1}^N C_j e^{-\vec r_i^2/R_j^2}\right)
d^{3}r_i
\nonumber\\                                   
          &=& \rho_c \int d^3 {\alpha} \left(\sum_{j=1}^N C_j
e^{i (\vec q+\vec \alpha) \cdot \vec r_1}
e^{-\vec r_1^2/R_j^2}\right) d^3 r_1
\prod_{i=2}^4 \left(\sum_{j=1}^N C_j
e^{-\vec r_i^2/R_j^2}
e^{i \vec \alpha \cdot \vec r_i}\right) d^3 r_i.
\label{eq12}
\end{eqnarray}                                
Integrating it with respect to $r_i$ we obtain
\begin{eqnarray}                              
F(\vec q)&=& \rho_c \int d^3 \alpha
\left(\sum_{j=1}^N C_j \left(\pi R_j^2\right)^{3/2}
e^{ \frac{R_j^2}{4} (\vec q+\vec \alpha)^2}\right)
\prod_{i=2}^4 \left(\sum_{j=1}^N C_j \left(\pi R_j^2\right)^{3/2}
e^{- \frac{R_j^2 \vec \alpha^2}{4}}\right)
\\                                   
    &=& \rho_c \int d^3 \alpha
\left(\sum_{j=1}^N C_j \left(\pi R_j^2\right)^{3/2}
e^{ \frac{R_j^2}{4} (\vec q+\vec \alpha)^2}\right)
\left(\sum_{j=1}^N C_j \left(\pi R_j^2\right)^{3/2}
e^{- \frac{R_j^2 \vec \alpha^2}{4}}\right)^3
\nonumber\\                                   
   &=& \rho_c \sum_{i_1,i_2,i_3,i_4=1}^N C_{i_1}C_{i_2}C_{i_3}C_{i_4}
\int d^3 \alpha \left(\prod_{j=1}^4 \left(\pi
R_{i_j}^2\right)^{3/2}\right)
e^{ \frac{R_{i_1}^2}{4} (\vec q^2+2 \vec \alpha \cdot \vec q)}
e^{- \frac{ \alpha^2}{4} \left(\sum_{j=1}^4 R_{i_j}^2\right)}
\nonumber\\                                   
  &=& \rho_c \sum_{i_1,i_2,i_3,i_4=1}^N C_{i_1}C_{i_2}C_{i_3}C_{i_4}
\left(\prod_{i_j=1}^A \left(\pi R_{i_j}^2\right)^{3/2}\right)
\left(\frac{4 \pi }{\sum_{j=1}^4 R_{i_j}^2}\right)^{3/2}\cdot
\nonumber\\
 & & \cdot exp\left( -
\frac{R_1^2 \vec q^2}{4}\left[\frac{\sum_{j=2}^4R_{i_j}^2}.
{\sum_{i_j=1}^4 R_{i_j}^2}\right]\right).\nonumber\\
\nonumber
\label{eq13}
\end{eqnarray}                              

The charge form factor, $F_{ch}(\vec q)$, of the ${}^4$He is connected
with $F(\vec q)$,
\begin{equation}
F_{ch}(\vec q)= F(\vec q) G_N (\vec q),
\label{eq14}
\end{equation}                             
where $G_N(\vec q)$  is the nucleon charge form factor,
$G_N(\vec q)=G_p(\vec q)+G_n(\vec q)$.  $G_p$ is the proton form factor
chosen in the dipole form \cite{W2}, $G_p(t)=(1-t/0.71)^{-2}$.  $G_n$
is the neutron form factor, $G_n(\vec q)=(1+r^2_1 q^2)^{-2}-(1+r^2_2
q^2)^{-2}$, where $r^2_1=1.24~(GeV/c)^{-2}$,$r^2_2=1.50~(GeV/c)^{-2}$
\cite{Nik-Dakhno}.  $t=-q^2$ is the four momentum transfer in
$(GeV/c)^2$.

In Fig. 2 the charge form factor calculations at the different
parametrizations ($B-D$) are compared with the experimental data of
R.F.  Frosch et. al.  \cite{Frosch}.
The charge form factor predicted by parametrization $A$ is not
presented because it does not reproduce the data at $q^2>0.35$ $(GeV/c)^2$.
As seen, at small values of $t$ all parametrizations give the same good
description of the data. They are different only at large values of $t$
due to the difference of the corresponding one-particles densities in
the center of the nucleus (see Fig. 1). We consider parametrization $D$
as the best one though it gives a dip position at a somewhat
smaller value of $t$ than it is needed for a perfect description of the
data. We think that the inclusion of the twelve quark bag component of
the ground state wave function will not change the results drastically
(see consideration in Ref. \cite{Nik-Dakhno}).
\newpage
\begin{figure}[t]
\begin{center}
\psfig{file=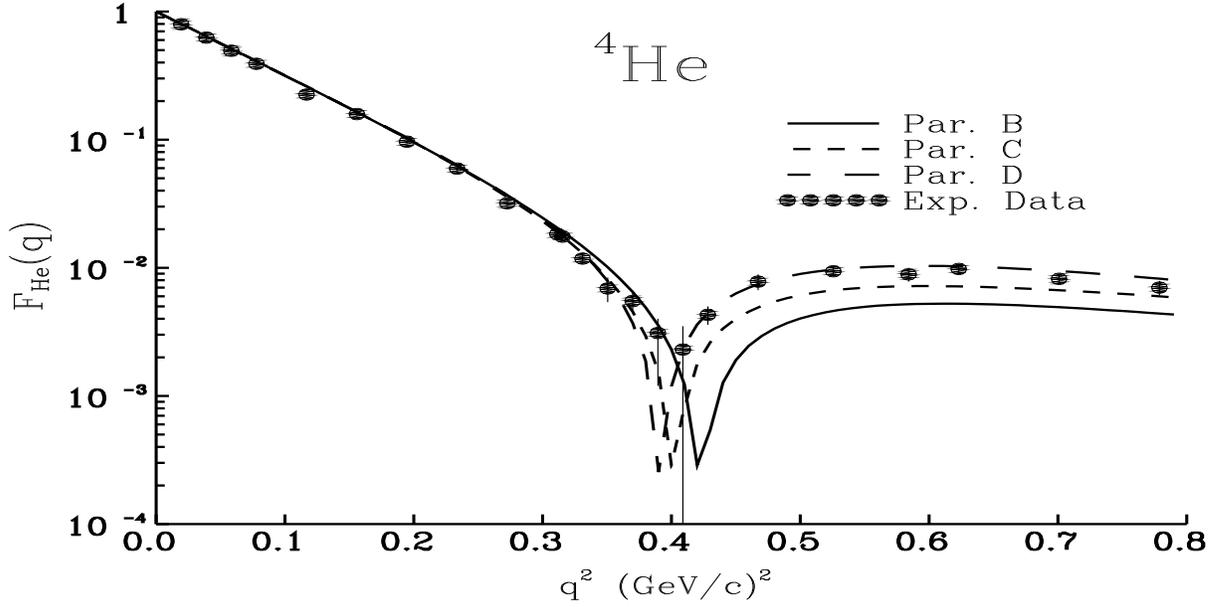,width=160mm,height=80mm,angle=0}
\caption{The charge form-factor of the ${}^4He$ nucleus.
The points are the experimental data \protect{\cite{Frosch}}, lines --
our calculations.}
\end{center}
\end{figure}

\section{The differential elastic cross section}

The Glauber amplitude for hadron-nucleus
scattering has a form \cite{Glauber}:
\begin{equation}                                 
F_{1A}(\vec q)=\frac{ip}{2\pi}\int d^2b~ e^{i\vec q \cdot
\vec b}\langle\psi_f |1-\prod_{j=1}^{A}\left(1-\gamma(\vec b - \vec
s_j)\right)|\psi_i\rangle ,
\label{eq15}
\end{equation}                                   
where $\vec b$ is the impact parameter, $p$ is the momentum of
the projectile hadron, $\psi_i$ and $\psi_f$ are initial and
final states wave functions, respectively. $\gamma$ is the $NN$ elastic
scattering amplitude in the impact parameter representation.
The corresponding differential cross section is given as
\begin{equation}                                 
\frac{d\sigma}{d\Omega} = \left|F_{1A}\right|^2.
\label{eq16}
\end{equation}                                   
In the case of the elastic $p{}^4He$ scattering
the amplitude $F_{1A}$ given by Eq. (\ref{eq15}) can be re-written as
\begin{equation}                                 
F_{14}(\vec q)~=~\frac{ip}{2\pi} \int d^2b~ e^{i \vec q \cdot \vec b}
\left[1-\prod_{j=1}^4 \left(1-\gamma(\vec b-\vec s_j) \right)\right]
\left|\psi(\vec r_1,\ldots,\vec r_A) \right|^2 \prod_{j=1}^4 d^2 r_j.
\label{eq17}
\end{equation}                                    
Substituting Eq. (\ref{eq2}) in Eq. (\ref{eq17}) gives
\begin{equation}                                  
F_{14}(\vec q)~=~\frac{ip}{2\pi} \rho_c \int d^2b~d^3\alpha
e^{i \vec q \cdot \vec b} \left[1-\prod_{j=1}^4 \left(1-\gamma(\vec
b-\vec s_j) \right)\right] e^{i \vec \alpha \cdot \sum_{j=1}^4 \vec r_j
} \prod_{j=1}^4 \phi(\vec r_j)d^3 r_j,
\label{eq18}
\end{equation}                                    
\noindent where $\vec r=\vec s +\vec z, \vec z$ is the component of the
position vector $\vec r$ in the direction along the projectile
momentum $\vec p$. We assume it is the direction of the z-axis.
Taking into account the relation (\ref{eq4}), we obtain
\begin{equation}                                  
\frac{2\pi}{ip\rho_c}F_{14}(\vec q)=
\int d^2b~ e^{i \vec q \cdot \vec b} \left[1-\prod_{j=1}^4
\left(1-\gamma(\vec b-\vec s_j) \right)\right]
\left(d^3\alpha ~e^{i \vec \alpha \cdot
\left( \sum_{j=1}^4 \vec r_j \right) }\right)\cdot
\label{eq19}
\end{equation}                                   
$$
\cdot \prod_{j=1}^4\sum_{k=1}^N C_k
  e^{-\vec r_j^2/R_k^2}d^3 r_j
=\int d^2b~d^3\alpha e^{i \vec
q \cdot \vec b} \left[1-\prod_{j=1}^4 \left(1-\gamma(\vec b-\vec s_j)
\right)\right] \cdot
$$
$$
\cdot
\prod_{j=1}^4\sum_{k=1}^N C_k~e^{i \vec \alpha \cdot
\vec r_j} e^{-\vec r_j^2/R_k^2} d^3 r_j,
$$
\noindent $\vec \alpha=\vec \alpha_2 +\vec \alpha_1$. As before $ \vec
\alpha_1 $ is the component of $\vec \alpha$ in the z-axis direction,
and $\vec \alpha_2$ is two-dimensional vector orthogonal to $\vec p$.
Having that
\begin{eqnarray}                                 
1-\prod_{j=1}^4 \left(1-\gamma(\vec b- \vec s_j)\right)&=&
\sum_{k=1}^4\gamma(\vec b-\vec s_k)-\sum_{k_1,k_2=1}^4\gamma(\vec b
-\vec s_{k_1}) \gamma(\vec b-\vec s_{k_2})
\nonumber\\
 & & +\sum_{k_1,k_2,k_3=1}^4 \gamma(\vec b-\vec s_{k_1})\gamma(\vec
b-\vec s_{k_2})\gamma(\vec b-\vec s_{k_3})
\nonumber\\
& & - \gamma(\vec b-\vec s_{k_1})\gamma(\vec b-\vec s_{k_2})\gamma(\vec
b-\vec s_{k_3})\gamma(\vec b-\vec s_{k_4}),
\label{20}
\end{eqnarray}                                   
\noindent we can write the amplitude as a sum of the multiple
scattering terms
\begin{equation}                                  
\frac{2\pi}{ip\rho_c}F_{14}(\vec  q) ~=~F_{14}^{(1)} - F_{14}^{(2)} +
F_{14}^{(3)} - F_{14}^{(4)}.
\label{eq21}
\end{equation}                                    
\noindent Every term can be calculated separately if $\gamma$ is chosen
as
\begin{equation}                                  
\gamma(\vec b)=\beta~e^{-\vec b^2/2B_{NN}},
\label{eq_gamma}
\end{equation}                                    
\noindent where $\beta= \left(\sigma_{NN}^{tot}(1-i\alpha_{NN}\right)
/\left(4\pi B_{NN}\right)$, $\sigma_{NN}^{tot}$ is the $NN$ total cross
section, $B_{NN}$ -- the slope parameter of the $NN$ differential
elastic cross section at zero momentum transfer, $\alpha_{NN}$ -- the
ratio of the real to imaginary parts of the $NN$ elastic scattering
amplitude at zero momentum transfer.  Then, the first term will be
\begin{eqnarray}                                    
F_{14}^{(1)}&=&\int d^2b~d^3\alpha
\left(\sum_{k_1=1}^4 \gamma(\vec b-\vec s_{k_1})\right) e^{i \vec q
\cdot \vec b} ~~~~\prod_{j=1}^4\sum_{i=1}^N C_i~e^{i \vec \alpha \cdot
\vec r_j} e^{-r_j^2/R_i^2} d^3 r_j
\nonumber\\                                        
&=&\int d^2b~d^3\alpha e^{i \vec q \cdot \vec b}
\left( \sum_{k_1=1}^4 \gamma(b-s_{k_1}) \right)
\left( \sum_{i=1}^N C_i~e^{i \vec \alpha \cdot \vec r_{k_1}}
e^{-r_{k_1}^2/R_i^2} d^3 r_{k_1} \right)
\nonumber\\                                        
 & & \prod_{k_2,k_3,k_4}
\left(\sum_{i=1}^N C_i~e^{i \vec \alpha \cdot \vec r_{k_j}}
e^{-r_{k_j}^2/R_i^2} d^3 r_{k_j} \right)
\nonumber\\                                        
&=& \sum_{k_1=1}^4\int d^2b~d^3\alpha
\gamma(b-s_{k_1}) e^{i \vec q \cdot \vec b} \left(\sum_{i=1}^N
C_i~e^{i \vec \alpha \cdot \vec r_{k_1}} e^{-r_{k_1}^2/R_i^2} d^3
r_{k_1}\right)
\nonumber\\                                         
 & & \prod_{k_2,k_3,k_4} \left(\sum_{i=1}^N C_i~e^{i \vec \alpha
\cdot \vec r_{k_j}} e^{-r_{k_j}^2/R_i^2} d^3 r_{k_j}\right)
\nonumber\\                                         
&=& \sum_{k_1=1}^4\int d^2b~d^3\alpha e^{i \vec q \cdot
\vec b} \left(\sum_{i=1}^N C_i \gamma(b-s_{k_1}) ~e^{i
\vec \alpha \cdot \vec r_{k_1}} e^{-r_{k_1}^2/R_i^2} d^3
r_{k_1}\right)
\nonumber\\                                          
 & & \prod_{k_2,k_3,k_4} \left(\sum_{i=1}^N C_i~e^{i \vec
\alpha \cdot \vec r_{k_j}} e^{-r_{k_j}^2/R_i^2} d^3 r_{k_j}\right)
\nonumber\\                                         
&=&\sum_{k_1=1}^4\beta\int d^2b~d^3\alpha e^{i \vec q \cdot
\vec b} \left(\sum_{i=1}^N C_i e^{\left(b-s_{k_1}\right)^2}~e^{i \vec
\alpha \cdot \vec r_{k_1}} e^{-r_{k_1}^2/R_i^2} d^3 r_{k_1}\right)
\nonumber\\                                         
& & \prod_{k_2,k_3,k_4} \left(\sum_{i=1}^N C_i~\left( \pi R_i^2
\right)^{3/2} e^{-\alpha^2 R_i^2 /2}\right)
\nonumber\\                                        
&=&\sum_{k_1=1}^4\beta\int d^2b~d^3\alpha e^{i \vec q \cdot
\vec b} \left(\sum_{i=1}^N C_i~\left( \pi R_i^2 \right)^{3/2}
e^{-\alpha^2 R_i^2 /2}\right)^3
\nonumber\\                                        
&&e^{\frac{1}{4}\left(\frac{2BR_i^2}{2B+R_i^2}\right)
\left(\frac{b}{B}+i\alpha_2 \right)^2}
\left(\sum_{i=1}^N C_i e^{-b^2/2B} e^{-\alpha_1^2
R_i^2/4} \left((\pi R_i^2 \right)^{1/2}\right.
\nonumber\\                                        
& & \left.exp\left[-\left(\frac{1}{R_i^2}+\frac{1}{2B}\right)
\left(s_{k_1}^2 -\frac{1}{2}\left(\frac{b}{B}+i\alpha_2 \right) \left(
\frac{2BR_i^2}{2B+R_i^2}\right) \right)^2 \right]\right)
\nonumber\\                                         
&=&\sum_{k_1=1}^4\beta\int d^2b~d^3\alpha e^{i \vec q
\cdot \vec b} \left(\sum_{i=1}^N C_i~\left( \pi R_i^2 \right)^{3/2}
e^{-\alpha^2 R_i^2 /2}\right)^3
\nonumber\\                                         
& & \left(\sum_{i=1}^N C_i e^{-b^2/2B} e^{-\alpha_1^2
R_i^2/4} \left(\pi R_i^2 \right)^{1/2} \left(\frac{2 \pi B R_i^2}
{2B+R_i^2} \right)\right.
\nonumber\\                                         
&&\left. exp\left[\frac{2B}{4}\left(\frac{R_i^2}{2B+R_i^2}\right)
\left(\frac{b}{B}+i\alpha_2 \right)^2\right]\right)
\nonumber\\                                         
&=&4\beta~\sum_{k_{i_j},j=1}^4 C_{k_1}C_{k_2}C_{k_3}C_{k_4}
\left(\prod_{j=1}^{4}\left( \pi R_{k_j}^2 \right)^{3/2}\right)
\left(\frac{2B}{2B+R_{k_1}^2}\right)
\int d^2b~d^3\alpha e^{i \vec q \cdot \vec b}
\nonumber\\                                         
& & e^{-\frac{\alpha_1^2}{4} \sum_{j=1}^{4}R_{k_j}^2}
e^{-\frac{\alpha_2^2}{4} \left(\sum_{j=2}^4R_{k_j}^2\right)} e^{-b/B}
e^{i\vec \alpha \cdot \vec b \left(
\frac{R_{k_1}^2}{R_{k_1}^2+2B}\right)} e^{-\frac{b^2}{2B}
\left(\frac{R_{k_1}^2}{R_{k_1}^2+2B}\right)} e^{-\frac{B
\alpha_2^2}{2} \left( \frac{R_{k_1}^2}{R_{k_1}^2+2B}\right)}
\nonumber\\                                         
&=&4\beta~\sum_{k_{i_j},j=1}^4 C_{k_1}C_{k_2}C_{k_3}C_{k_4}
\left(\prod_{j=1}^{4}\left( \pi R_{k_j}^2 \right)^{3/2}\right)
\left(\frac{2B}{2B+R_j^2}\right)
\left(\frac{4\pi}{\sum_{j=1}^4 R_{k_j}^2}\right)^{1/2}
\nonumber\\                                         
& & \int d^2b~d^3\alpha e^{-B \alpha_2^2 S_1/2}
e^{-\frac{\alpha_2^2}{4} \left(\sum_{j=2}^4R_{k_i}^2\right)} e^{-b/B}
e^{-\left(i(\vec q+\vec \alpha_2 S_1)/2M_1 \right)^2}
\nonumber\\                                        
& &exp\left[-M_1\left(b^2-i\left(\vec q+\vec \alpha_2 S_1\right)\cdot
\vec b/M_1+\left(i(\vec q+\vec \alpha_2
S_1)/2M_1\right)^2\right) \right]
\label{eq22}
\end{eqnarray}                                     
where
$$
M_1=\frac{1}{B}-\frac{1}{B}
\frac{R_{k_1}^2}{R_{k_1}^2+2B}=  \frac{1}{R_{k_1}^2+2B},~~~
S_1=\frac{R_{k_1}^2}{R_{k_1}^2+2B}.
$$
\begin{eqnarray}                                    
F_{14}^{(1)}&=&4\beta~\sum_{k_{i_j},j=1}^4 C_{k_1}C_{k_2}C_{k_3}C_{k_4}
\left(\prod_{j=1}^{4}\left( \pi R_{k_j}^2 \right)^{3/2}\right)
\left(\frac{2B}{2B+R_j^2}\right)^{3/2}
\left(\frac{4\pi}{\sum_{j=1}^4 R_{k_j}^2}\right)^{1/2}
\nonumber\\                                        
& & \left(\frac{\pi}{M_1}\right)
\left(\frac{4\pi}{H_1}\right)
e^{-q^2/4M_1} exp\left[q^2S_1^2/4M_1 H_1 \right]
\label{eq23}
\end{eqnarray}                                     
$$
H_1=R_{k_1}^2+R_{k_3}^2+R_{k_4}^2+2BS_1+\frac{S_1^2}{M_1}.
$$
The second term also will be,
\begin{eqnarray}                                    
F_{14}^{(2)}&=&\int d^2b~d^3\alpha
\left(\sum_{k_1,k_2=1}^4 \gamma(b-s_{k_1})\gamma(b-s_{k_2})\right) e^{i
\vec q \cdot \vec b}
~~~~\prod_{j=1}^4\sum_{i=1}^N C_j~e^{i \vec \alpha
\cdot \vec r_j} e^{-r_j^2/R_i^2} d^3 r_i
\nonumber\\                                        
&=&\int d^2b~d^3\alpha e^{i \vec q \cdot \vec b}
\left(\sum_{k_1,k_2=1}^4 \gamma(b-s_{k_1})\gamma(b-s_{k_2})\right)
\prod_{k_1,k_2}\left(\sum_{i=1}^N C_i~e^{i \vec \alpha \cdot \vec
r_{k_j}} e^{-r_{k_j}^2/R_i^2} d^3 r_{k_j}\right)
\nonumber\\                                        
 & & \prod_{k_3,k_4}
\left(\sum_{i=1}^N C_i~e^{i \vec \alpha \cdot \vec r_{k_j}}
e^{-r_{k_j}^2/R_i^2} d^3 r_{k_j}\right)
\nonumber\\                                        
&=& \sum_{k_1,k_2=1}^4\int d^2b~d^3\alpha
\gamma(b-s_{k_1})\gamma(b-s_{k_2})
e^{i \vec q \cdot \vec b}
\prod_{k_1,k_2}\left(\sum_{i=1}^N
C_i~e^{i \vec \alpha \cdot \vec r_{k_j}} e^{-r_{k_j}^2/R_i^2} d^3
r_{k_j}\right)
\nonumber\\                                         
 & & \prod_{k_3,k_4} \left(\sum_{i=1}^N C_i~e^{i \vec \alpha
\cdot \vec r_{k_j}} e^{-r_{k_j}^2/R_i^2} d^3 r_{k_j}\right)
\nonumber\\                                         
&=& \sum_{k_1,k_2=1}^4\int d^2b~d^3\alpha e^{i \vec q \cdot
\vec b} \prod_{k_1,k_2}\left(\sum_{i=1}^N C_i \gamma(b-s_{k_j}) ~e^{i
\vec \alpha \cdot \vec r_{k_j}} e^{-r_{k_j}^2/R_i^2} d^3
r_{k_j}\right)
\nonumber\\                                          
 & & \prod_{k_3,k_4} \left(\sum_{i=1}^N C_i~e^{i \vec
\alpha \cdot \vec r_{k_j}} e^{-r_{k_j}^2/R_i^2} d^3 r_{k_j}\right)
\nonumber\\                                         
&=&\sum_{k_1,k_2=1}^4\beta^2\int d^2b~d^3\alpha e^{i \vec q \cdot
\vec b} \prod_{k_1,k_2}\left(\sum_{i=1}^N C_i
e^{\left(b-s_{k_j}\right)^2}~e^{i \vec \alpha \cdot
\vec r_{k_j}} e^{-r_{k_j}^2/R_i^2} d^3 r_{k_j}\right)
\nonumber\\                                         
 & & \prod_{k_3,k_4} \left(\sum_{i=1}^N C_i~\left( \pi R_i^2
\right)^{3/2} e^{-\alpha^2 R_i^2 /2}\right)
\nonumber\\                                        
&=&\sum_{k_1,k_2=1}^4\beta^2\int d^2b~d^3\alpha e^{i \vec q \cdot
\vec b} \left(\sum_{i=1}^N C_i~\left( \pi R_i^2 \right)^{3/2}
e^{-\alpha^2 R_i^2 /2}\right)^2
\nonumber\\                                        
& & \prod_{k_1,k_2}\left(\sum_{i=1}^N C_i e^{-b^2/2B} e^{-\alpha_1^2
R_i^2/4} \left((\pi R_i^2 \right)^{1/2}\right.
e^{\frac{1}{4}\left(\frac{2BR_i^2}{2B+R_i^2}\right)
\left(\frac{b}{B}+i\alpha_2 \right)^2}
\nonumber\\                                        
& & \left.exp\left[-\left(\frac{1}{R_i^2}+\frac{1}{2B}\right)
\left(s_{k_j}^2 -\frac{1}{2}\left(\frac{b}{B}+i\alpha_2 \right) \left(
\frac{2BR_i^2}{2B+R_i^2}\right) \right)^2 \right]\right)
\nonumber\\                                         
&=&\sum_{k_1,k_2=1}^4\beta^2\int d^2b~d^3\alpha e^{i \vec q
\cdot \vec b} \left(\sum_{i=1}^N C_i~\left( \pi R_j^2 \right)^{3/2}
e^{-\alpha^2 R_j^2 /2}\right)^2
\nonumber\\                                         
& &\prod_{k_1,k_2}\left(\sum_{i=1}^N C_i e^{-b^2/2B} e^{-\alpha_1^2
R_i^2/4} \left(\pi R_j^2 \right)^{1/2} \left(\frac{2 \pi B R_j^2}
{2B+R_i^2} \right)\right.
\nonumber\\                                         
& & \left.exp\left[\frac{2B}{4}\left(\frac{R_i^2}{2B+R_i^2}\right)
\left(\frac{b}{B}+i\alpha_2 \right)^2\right]\right)
\nonumber\\                                         
&=&6\beta^2~\sum_{k_{i_j},j=1}^4 C_{k_1}C_{k_2}C_{k_3}C_{k_4}
\left(\prod_{j=1}^{4}\left( \pi R_{k_i}^2 \right)^{3/2}\right)
\prod_{j=1}^{2}\left(\frac{2B}{2B+R_i^2}\right)
\nonumber\\                                         
& &\int d^2b~d^3\alpha e^{i \vec q \cdot \vec b}
e^{-\frac{\alpha_1^2}{4} \sum_{j=1}^{4}R_{k_j}^2}
e^{-\frac{\alpha_2^2}{4} \left(R_{k_3}^2+R_{k_3}^2\right)}
\nonumber\\                                         
& & e^{-b/B}
e^{i\vec \alpha \cdot \vec b \left(\sum_{j=1}^{2}
\frac{R_{k_j}^2}{R_{k_j}^2+2B}\right)}
e^{-\frac{b^2}{2B} \left(\sum_{j=1}^{2}
\frac{R_{k_j}^2}{R_{k_j}^2+2B}\right)}
e^{-\frac{B \alpha_2^2}{2} \left(\sum_{j=1}^{2}
\frac{R_{k_j}^2}{R_{k_j}^2+2B}\right)}
\nonumber\\                                         
&=&6\beta^2~\sum_{k_{i_j},j=1}^4 C_{k_1}C_{k_2}C_{k_3}C_{k_4}
\left(\prod_{j=1}^{4}\left( \pi R_{k_j}^2 \right)^{3/2}\right)
\prod_{j=1}^{2}\left(\frac{2B}{2B+R_j^2}\right)
\left(\frac{4\pi}{\sum_{j=1}^4 R_{k_j}^2}\right)^{1/2}
\nonumber\\                                         
& & \int d^2b~d^3\alpha e^{-B \alpha_2^2 S_2/2}
e^{-\frac{\alpha_2^2}{4} \left(R_{k_3}^2+R_{k_4}^2\right)}e^{-b/B}
e^{-\left(i(\vec q+\vec \alpha_2 S_2)/2M_2 \right)^2}
\nonumber\\                                        
& &exp\left[-M_2\left(b^2-i\left(\vec q+\vec \alpha_2 S_2\right)\cdot
\vec b/M_2+\left(i(\vec q+\vec \alpha_2
S_2)/2M_2\right)^2\right) \right],
\label{eq24}
\end{eqnarray}                                     
where
$$
M_2=\frac{1}{B}-\frac{1}{2B} \sum_{j=1}^2
\frac{R_{k_j}^2}{R_{k_j}^2+2B}= \sum_{j=1}^2 \frac{1}{R_{k_j}^2+2B},~~~
S_2=\sum_{j=1}^2 \frac{R_{k_j}^2}{R_{k_j}^2+2B}.
$$
\begin{eqnarray}                                    
F_{14}^{(2)}&=&6\beta^2~\sum_{k_{i_j},j=1}^4 C_{k_1}C_{k_2}C_{k_3}C_{k_4}
\left(\prod_{j=1}^{4}\left( \pi R_{k_j}^2 \right)^{3/2}\right)
\prod_{j=1}^{2}\left(\frac{2B}{2B+R_j^2}\right)
\left(\frac{4\pi}{\sum_{j=1}^4 R_{k_j}^2}\right)^{1/2}
\nonumber\\                                        
& & \left(\frac{\pi}{M_2}\right)
\left(\frac{4\pi}{H_2}\right)
e^{-q^2/4M_2} exp\left[q^2S_2^2/4M_2 H_2 \right]
\label{eq25}
\end{eqnarray}                                     
$$
H_2=R_{k_3}^2+R_{k_4}^2+2BS_2+\frac{S_2^2}{M_2}.
$$
By the
same way the other terms will be
\begin{eqnarray}                                    
F_{14}^{(3)}&=&4\beta^3~\sum_{k_{i_j},j=1}^4 C_{k_1}C_{k_2}C_{k_3}C_{k_4}
\left(\prod_{j=1}^{4}\left( \pi R_{k_j}^2 \right)^{3/2}\right)
\prod_{j=1}^{2}\left(\frac{2B}{2B+R_j^2}\right)
\left(\frac{4\pi}{\sum_{j=1}^4 R_{k_j}^2}\right)^{1/2}
\nonumber\\                                         
& & \left(\frac{\pi}{M_3}\right)
\left(\frac{4\pi}{H_3}\right)
e^{-q^2/4M_3} exp\left[q^2S_3^2/4M_3 H_3 \right],
\label{eq26}
\end{eqnarray}                                      
$$
M_3=\frac{3}{2B}-\frac{1}{2B} \sum_{j=1}^3
\frac{R_{k_j}^2}{R_{k_j}^2+2B} = \sum_{j=1}^3 \frac{1}{R_{k_j}^2+2B}
,~~~~ S_3=\sum_{j=1}^3 \frac{R_{k_j}^2}{R_{k_j}^2+2B}, ~~~~
$$
$$H_3=R_{k_4}^2+2BS_3+\frac{S_3^2}{M_3}
$$
\begin{eqnarray}                                 
F_{14}^{(4)}&=&\beta^4~\sum_{k_{i_j},j=1}^4 C_{k_1}C_{k_2}C_{k_3}C_{k_4}
\left(\prod_{j=1}^{4}\left( \pi R_{k_j}^2 \right)^{3/2}\right)
\prod_{j=1}^{2}\left(\frac{2B}{2B+R_j^2}\right)
\left(\frac{4\pi}{\sum_{j=1}^4 R_{k_j}^2}\right)^{1/2}
\nonumber\\                                    
& & \left(\frac{\pi}{M_4}\right)
\left(\frac{4\pi}{H_4}\right)
e^{-q^2/4M_4} exp\left[q^2S_4^2/4M_4 H_4 \right]
\label{eq27}
\end{eqnarray}                                
$$
M_4=\frac{2}{B}-\frac{1}{2B}
\sum_{j=1}^4 \frac{R_{k_j}^2}{R_{k_j}^2+2B}=
\sum_{j=1}^4 \frac{1}{R_{k_j}^2+2B},~~~
S_4=\sum_{j=1}^4 \frac{R_{k_j}^2}{R_{k_j}^2+2B},~~~
H_4=2BS_4+\frac{S_4^2}{M_4}.
$$

In many experimental papers \cite{Velichko,Bujak,Burq} the authors
included the Coulomb scattering amplitude in a simple way in order
to extract the nuclear total $p{}^4He$ cross section,
\begin{equation}                              
\frac{d\sigma}{dt} = \left|F_ce^{i\phi}+F_N\right|^2,
\label{eq28}
\end{equation}                                
\noindent where
\begin{equation}                             
F_c(t)=\frac{4\alpha \sqrt{\pi}}{\beta t}G_p(t)G_{He}(t),
\label{29}
\end{equation}                               
\noindent $\alpha=1/137$ is the fine structure constant, $\beta= v/c$
is the proton velocity in the laboratory system, $G_p(t),G_{He}(t)$ are
the electromagnetic form factor of the proton and $He$, respectively,
\begin{equation}                            
G_{He}(t)=exp\left[\frac{r_{He}^2 t}{6}\right],
\label{eq30}
\end{equation}                              
$$r_{He}^2=r_e^2-r_p^2,~~~ r_e=1.67~ fm,~~~ r_p=0.812~ fm.$$
\begin{equation}                            
\phi=2\alpha \beta^{-1}\left[ln\left(B|t|\right)+0.577...\right].
\label{eq31}
\end{equation}                              
$B=29 \left(GeV/c\right)^{-2}$ and $t=-q^2$.
$F_n$ is the nuclear amplitude written at small $t$ in the form
\begin{equation}                            
F_n=\frac{\sigma^{tot}}{4\hbar\sqrt{\pi}}\left(i+\alpha\right)
e^{\frac{1}{2}Bt}.
\label{eq32}
\end{equation}                              
We follow the same way replacing $F_N$ by the Glauber
scattering amplitude.

The available experimental data on the $p{}^4He$ elastic scattering have
been presented by G.N.~Velichko et. al. \cite{Velichko} at the
energies of 0.695, 0.793, 0.89, 0.991 $GeV$; by A. Bujak et. al.
\cite{Bujak} at the energies of 45, 97, 146, 200, 259, 301, 393
$GeV$, and by J.P. Burq et al.  \cite{Burq} at the energies of 100,
150, 250, 300 $GeV$. To calculate the Glauber amplitudes
at these energies, it is needed to have the values of the
nucleon-nucleon amplitude parameters $\sigma_{NN}^{tot},~ B_{NN}$ and
$\alpha_{NN}$.  $\sigma_{NN}^{tot}$ was estimated as an average of the
neutron-proton total cross section, $\sigma_{tot}^{np}$, and
the proton-proton total cross section, $\sigma_{tot}^{pp}$, which can be
taken from the compilation of the experiential data \cite{CERN}.

More complicated situation is with $B_{NN}$. There are only few
experimental data, and it is not enough for all energies.  Thus we
have used another way to evaluate $B_{NN}$ from the total and
elastic $NN$ cross sections. At chosen form of $\gamma (\vec b)$ (see
Eq. \ref{eq_gamma}) the elastic $NN$ cross section, $\sigma^{el}_{NN}$,
is given as
\begin{eqnarray}          
\sigma^{el}_{NN}&=&
\frac{1}{p^2}\left(\frac{p}{2\pi}\right)^2\int \gamma(\vec b_1)
e^{i\vec q \cdot \vec b_1} \gamma^*(\vec b_2) e^{-i\vec q \cdot \vec
b_2} d^2b_1 d^2b_2 d^2q \nonumber\\
&=& \frac{1}{\left(2\pi\right)^2}\int e^{i\vec q
\cdot (\vec b_1-\vec b_2} \gamma(\vec b_1) \gamma^*(\vec b_2) d^2b_1
d^2b_2 d^2q \nonumber\\
&=& \int \delta(\vec b_1-\vec b_2) \gamma(\vec b_1)
\gamma^*(\vec b_2) d^2b_1 d^2b_2
=\int \left| \gamma(\vec b) \right|^2 d^2 b
\nonumber\\                                   
&=& \left(\frac{\sigma^{tot}_{NN}}{4\pi B_{NN}}\right)^2 \left(1+
\alpha_{NN}^2 \right) \int e^{-\frac{b^2}{B_{NN}}} d^2b
\nonumber\\
&=& \left(\frac{\sigma^{tot}_{NN}}{4\pi B_{NN}}\right)^2 \left(1+
\alpha_{NN}^2 \right) \pi B_{NN}
= \frac{\left(\sigma^{tot}_{NN}\right)^2}{16 \pi B_{NN}}
\left(1+ \alpha_{NN}^2 \right).
\label{eq35}
\end{eqnarray}                                
Since $\alpha_{NN}$ is very small, we neglect it in our
calculations.  In this case $B_{NN}$ can be calculated as
\begin{equation}                            
B_{NN}=\frac{\left(\sigma_{NN}^{tot}\right)^2}{16\pi \sigma_{NN}^{el}}.
\label{36}
\end{equation}                              
$\sigma_{NN}^{el}$ was taken from the compilation of the experimental
data \cite{CERN} as an average of $pn$ and $pp$ cross sections.

The values of $\alpha_{NN}$ for all mentioned above energies were
extracted from the compilation of the experiential data \cite{UCRL}.
\begin{table}[h]                        
\centering
\caption{ The parameters used at the calculations of the Glauber
amplitudes}

\begin{tabular}{|c|c|c|c|c|c|c|} \hline
   $E_{kin}$ &$\sigma_{NN}^{el}$& $\sigma_{pp}^{tot}$ &
   $\sigma_{pn}^{tot}$ & $\sigma_{NN}^{tot}$ & $B_{NN}$ &$\alpha_{NN}$ \\
$GeV$ & $mb$  & $mb$  & $mb$  & $mb$  & $(GeV/c)^{-2}$ & \\ \hline
0.695 & 24.2  & 42.4  & 38.38 & 40.39 & 4.069 & -0.205\\ \hline
0.795 & 22.5  & 46.8  & 38.56 & 42.68 & 4.134 & -0.1975\\ \hline
0.890 & 24.4  & 47.3  & 38.73 & 43.01 & 3.872 & -0.19   \\ \hline
0.991 & 24.27 & 47.6  & 39.24 & 43.42 & 3.967 & -0.185\\ \hline
0.992 & 24.27 & 47.6  & 39.24 & 43.42 & 3.967 & -0.185\\ \hline
45    & 7.402 & 38.48 & 38.32 & 38.4  & 10.173& -0.087\\ \hline
97    & 6.985 & 37.94 & 38.89 & 38.4  & 10.783& -0.090\\ \hline
100   & 6.985 & 37.94 & 38.89 & 38.4  & 10.783& 0.1\\ \hline
146   & 7.03  & 38.29 & 39.12 & 38.71 & 10.884& -0.049\\ \hline
150   & 7.03  & 38.69 & 39.12 & 38.91 & 10.996& 0.105\\ \hline
200   & 6.895 & 38.98 & 39.56 & 39.27 & 11.422& -0.022\\ \hline
250   & 6.89  & 39.34 & 39.83 & 39.58 & 11.614& 0.11\\ \hline
259   & 6.89  & 39.34 & 39.83 & 39.58 & 11.614& 0.024\\ \hline
300   & 6.888 & 39.46 & 39.83 & 39.65 & 11.653& 0.115\\ \hline
301   & 6.888 & 39.46 & 39.83 & 39.65 & 11.653& 0.031\\ \hline
393   & 7.016 & 40.19 & 40.01 & 40.1  & 11.703& 0.067\\ \hline
\end{tabular} \end{table}                      

All the parameter values used for our calculation are presented in
Table 2. Typical results of the calculations in comparison with
experimental data \cite{Velichko,Bujak} are shown in Figs. 3, 4.
As seen, the model calculations are above the experimental
data. The first diffraction minimums are shifted to small
$t$. We can confirm now that the model calculations can not reproduce
the data with required accuracy. This pushed us to search for
modification of the model.

\section{The twelve quark bag admixture}
Any nucleus consists of $3A$-quarks. In the ground state the quarks
are forming clusters, bags and nucleons. Following \cite{Nik-Dakhno} we
assume that the ${}^4He$ wave function is given as
\begin{equation}                            
|q_1...q_{12}>=\alpha|NNNN>+\beta|12q>,~~~~~|\alpha|^2+|\beta|^2=1,
\label{eq37}
\end{equation}                              
where $|q_1...q_{12}>$ is the $12$- quark bound state wave function of
the ${}^4He$. $|12q>$ stands for a component left after projecting
$|q_1...q_{12}>$ onto the four nucleons state. Since $|NNNN>$
vanishes in the central part of ${}^4He$, $|12q>$ must strongly peak
in the central part of the ${}^4He$. According to the assumption
of Ref. \cite{Nik-Dakhno} $<NNNN|12q>=0$. Thus we neglect
$12q-NNNN$ interference terms that vanish at $q=0$ in the
elastic $p{}^4He$ scattering amplitude and write

\begin{equation}                                 
F_{14}=\left(1-w_{12q}\right)F_{Gl} + w_{12q}F_{12q},
\label{eq39}
\end{equation}                                   
where $F_{Gl}$ is the Glauber amplitude of the $p-4N$ scattering, and
$w_{12q}$ is the weight of the $12q$ bag quark state, $w_{12q}=|\beta|^2$.
We take the nucleon - twelve quark bag scattering
amplitude in a simple form,
\begin{equation}                                
F_{12q}=\frac{\sigma_{12q}}{2} e^{t b_{12q}/2},
\label{38}
\end{equation}                                  
where $\sigma_{12q}$ is the $N-12q$ bag total cross section,
and $b_{12q}$ is the slope parameter.

We found the parameters $\sigma_{12q}$, $b_{12q}$ and $w_{12q}$
fitting the experimental data \cite{Velichko,Bujak,Burq}. The values
are presented in Table 3. As one can see, the parameter uncertainty is
very large at low energies ($E_{kin} < 1~~GeV$). This means that at the
energies one does not need to add anything to the Glauber
amplitude.  At higher energies the values become more stable excepting
the results at 146 GeV.
\begin{table}[h]                            
\centering
\caption{ The fitting values of $w_{12q}$, $\sigma_{12q}$ and $b_{12q}$}
\begin{tabular}{|c|c|c|c|c|} \hline
   $E_{kin}$ & $w_{12q}$ & $\sigma_{12q}$ & $b_{12q}$ &$\chi^2/NOF$ \\
   $GeV$     &           &   $mb$         &$(GeV/c)^{-2}$& \\ \hline
0.695 & 4.1 $\pm$66.5& 126.0 $\pm$169.7& 32.6$\pm$100.5 &72/65\\
\hline
0.795 & 9.8  $\pm$1.8 & 117.0 $\pm$10.7& 40    $\pm$228.9  &45/81\\
\hline
0.890 &10.6 $\pm$174.4& 169.4 $\pm$464.4& 37.4$\pm$164.2 &56/95\\
\hline
45    &8.34 $\pm$0.92& 32.62  $\pm$10.23& 20.38$\pm$3.90  &557/127\\
\hline
97    &9.23 $\pm$1.31& 30.83  $\pm$13.37& 20.46$\pm$5.54  &146/84\\
\hline
146   &13.51$\pm$0.42& 65.57 $\pm$2.54& 32.46$\pm$0.52 &222/84\\
\hline
200   &10.51$\pm$1.83& 28.86 $\pm$17.24& 21.62 $\pm$7.33   &285/84\\
\hline
259   &9.72 $\pm$2.91& 29.10 $\pm$30.04& 22.11$\pm$12.69 &264/86\\
\hline
301   &11.08$\pm$1.28& 25.94 $\pm$11.78& 21.68$\pm$5.46  &173/86\\
\hline
393   &10.80$\pm$1.59 & 25.73 $\pm$14.95& 21.14$\pm$6.96  &118/85\\
\hline
\end{tabular} \end{table}                    

We have excluded from the fit the data at the energies of 100, 150,
250, and 300 $GeV$ \cite{Burq}. The data are above the Glauber
calculations. Thus at the fitting an unreasonable large weight of
12q-bag ($>~ 50$ \%) and $\sigma_{12q}$ was obtained. We believe
the data are not quite well normalized. To show this, we plot the data
at close energies \cite{Bujak} on the same figures 3, 4.

As seen, there is a clear difference between the two groups of
experimental data. Maybe, it is due to a normalization error. We do not
know a reason of the error. However, one can see that the data
by Ref.  \cite{Burq} are falling out from the whole set of the
experimental data, and it is not possible to fit them correctly.


The figures show influence of the 12q bag admixture on the
differential cross section. The inclusion of the admixture leads to
decreasing the Glauber amplitude if $\sigma_{12q}$ is smaller than
$\sigma_{pHe}$. In the region of the dip where the Glauber amplitude
vanishes, $F_{12q}$ is positive and shifts the dip to a larger values of
$t$. So, the hypothesis really allows one to solve the main part of the
problem.

\newpage
\begin{figure}[cbth]
\begin{center}
\psfig{file=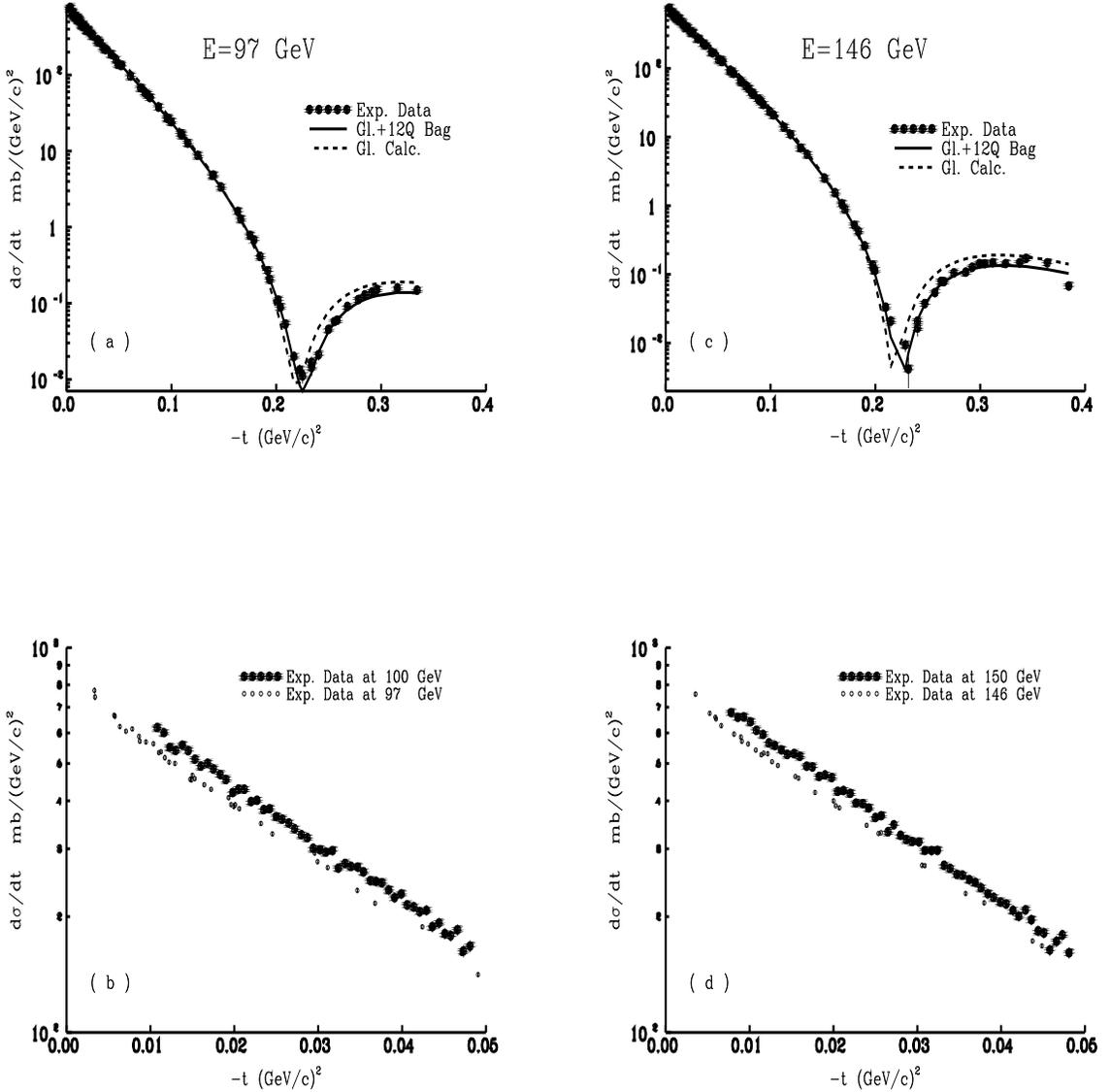,width=150mm,height=150mm,angle=-90}
\caption{The $p{}^4He$ differential elastic cross sections. The point
are the experimental data \protect\cite{Bujak,Burq}. The solid and
dashed lines are our calculations with and without 12q admixture,
respectively.}
\end{center}
\end{figure}
Clearly, inclusion of the inelastic screening into
calculations will lead to decreasing the cross section in the
region of small $t$, and to increasing in the region of the
large $t$ values. To compensate these, one have to
increase $w_{12q}$, $\sigma_{12q}$, and $b_{12q}$. From this point of
view we can understand the results of Ref. \cite{Nik-Dakhno}.
According to the Fig. 10 of the Ref. \cite{Nik-Dakhno}, $\sigma_{12q}
\sim 140~mb$ what is near to the $p{}4He$ total cross section, the
slope parameter of that $F_{12q}$ is larger than ours. As a result,
$w_{12q}\sim$ 12 \%. We have the average value of $w_{12q}\sim 10.5$\%
So, two values agree quite reasonable with each other. At the same
time, our $\sigma_{12q}$ is too small.

Let us mark that our amplitude $F_{12q}$ is more simple than that of
the Ref. \cite{Nik-Dakhno}. It can be easily used for future
calculations.

\newpage
\begin{figure}[cbth]
\begin{center}
\psfig{file=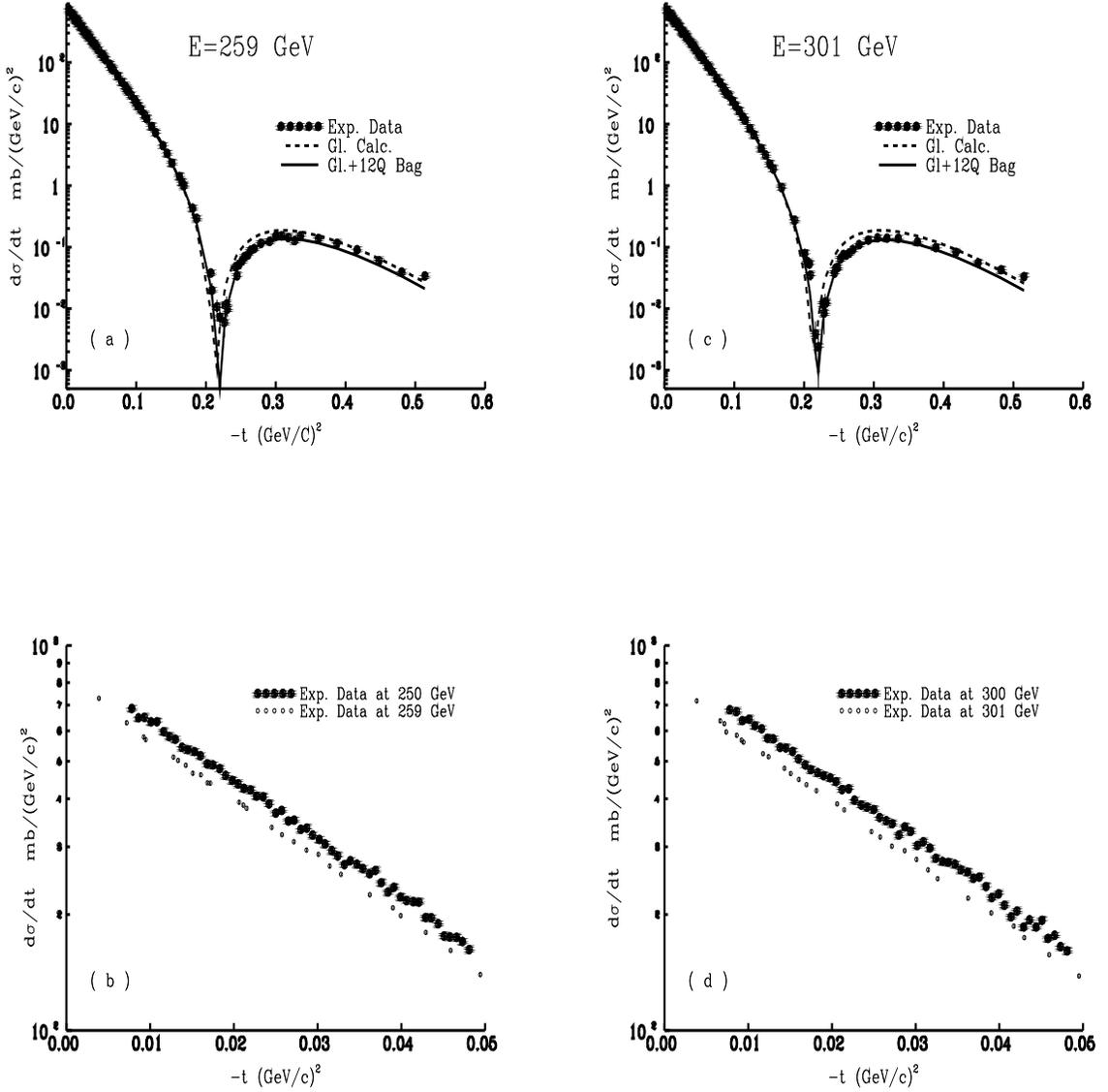,width=150mm,height=150mm,angle=-90}
\caption{The $p{}^4He$ differential elastic cross sections. Points show
experimental data \protect\cite{Bujak,Burq}. Solid and dashed lines are
our calculations with and without 12q admixture, respectively.}
\end{center}
\end{figure}

\section*{Conclusion}
The 12q bag admixture to the ground state wave function
of the ${}^4He$ nucleus allows one to describe quite well the elastic
$p{}^4He$ scattering. According to our estimations, the weight of the
12q bag is $\sim 10.5$\%, the proton - 12q bag total cross
section is $\sim 34$ mb, and the slope parameter
of the $p-12q$ bag elastic scattering is $\sim$ 23 (GeV/c)$^{-2}$.

V.V.Uzhinskii thanks RFBR (grand N 00-01-00307)
and INTAS (grand N 00-00366) for their financial support.
A.M. Mosallem thankfuls to Profs. A.B.B. Kalila and K.M. Hanna and for
support.
The authers also wishes to thank JINR officials for hospitality.

\end{document}